\documentclass[11pt,onecolumn,amssymb,nofootinbib]{revtex4}
\usepackage{amsmath, amsthm, amscd, amssymb}
\usepackage{bm}
\usepackage{bbm}

\begin{document}

\title{\bf Spacecraft calorimetry as a test of the dark matter scattering model \hfill\break for flyby anomalies}
\author{Stephen L. Adler}
\email{adler@ias.edu} \affiliation{Institute for Advanced Study,
Einstein Drive, Princeton, NJ 08540, USA.}

\begin{abstract}
In previous papers we have shown that
scattering of spacecraft nucleons from dark matter gravitationally bound to the earth
gives a possible explanation of the flyby velocity anomalies.  In addition to flyby
velocity changes arising from the average over the scattering cross section of the
collision-induced nucleon velocity change,  there will be spacecraft temperature increases
arising from the mean squared fluctuation of the collision-induced velocity change.
We give here a quantitative treatment of this effect, and suggest that careful calorimetry
on spacecraft traversing the region below 70,000 km where the flyby velocity changes take
place could verify, or at a minimum place significant constraints, on the dark matter scattering
model.

\end{abstract}

\maketitle

\section{Introduction}
In several recent papers we have explored the possibility that dark matter scattering may be
responsible for the anomalous geocentric frame
orbital energy changes that are observed during earth flybys of various spacecraft, as
reported by Anderson et al. \cite{anderson}.   Some flybys show energy decreases, and
others energy increases, with the largest anomalous velocity changes of order 1 part
in $10^6$.  While the possibility that these anomalies are artifacts of the orbital
fitting method used in \cite{anderson} is being actively studied, there is also
a chance that they may represent new physics.   In \cite{adler1} we explored, through order of magnitude estimates,
the possibility that the flyby anomalies
result from the scattering of spacecraft nucleons from dark matter particles in orbit around the earth, with the
observed velocity decreases arising from elastic scattering, and the observed velocity
increases arising from exothermic inelastic scattering, which can impart an energy impulse to a spacecraft nucleon.
In \cite{adler2} we constructed a concrete model, based on
two populations of dark matter particles, one of which scatters on nucleons elastically, and the other of which scatters inelastically, each with a shell-like distribution of  orbits generated by the precession of a
tilted circular orbit around the earth's rotation axis.  We showed in \cite{adler2} that this model can give a good fit to
the flyby data, with shell radii in the 30,000--35,000 km range.

In the present paper we follow up on the brief observation in \cite{adler1} that if  there is a spacecraft velocity change
as a result of dark matter scattering, there must be a corresponding temperature increase arising from fluctuations in
the scattering recoil direction.   In Sec. II we develop formulas for giving a quantitative treatment of this effect.  In Sec. III
we give order of magnitude quick estimates, and in Sec. IV we give numerical results based on the model of \cite{adler2}.
In Sec. V we suggest that a thermally shielded, spacecraft
based calorimetry experiment could potentially give crucial information on the dark matter scattering model for the flyby anomalies.

\section{Temperature change arising from velocity fluctuations}

In \cite{adler1} we considered the velocity change when a spacecraft nucleon of
mass $m_1\simeq 1 {\rm GeV}$ and initial velocity $\vec u_1$
scatters from a dark matter particle of mass $m_2$ and initial
velocity $\vec u_2$, into an outgoing nucleon of mass $m_1$ and
velocity $\vec v_1$, and an outgoing secondary dark matter
particle of mass $m_2'=m_2-\Delta m$ and velocity $\vec v_2$ . (In
the elastic scattering case, one has $m_2'=m_2$ and $\Delta m=0$.)
Under the assumption that  both initial particles are nonrelativistic, so that
$|\vec u_1|<<c, |\vec u_2|<<c$,  a
straightforward calculation shows that the outgoing nucleon velocity is given by
\begin{equation}\label{eq:vel1}
 \vec v_1=\frac {m_1 \vec u_1 + m_2 \vec
u_2}{m_1+m_2'}+w\hat v_{\rm out}~~~.
\end{equation}
Here $w>0$ is given\footnote{The notation $t$ was used in \cite{adler1} for what we here term $w$;
the change in notation avoids confusion with use of $t$ for time. We take $\hbar=1$ in Sec. III, while the velocity 
of light is denoted throughout by $c$.  } by taking the square root of
\begin{equation}\label{eq:tdef}
w^2=\frac{m_2m_2'}{(m_1+m_2)(m_1+m_2')}(\vec u_1-\vec u_2)^2 +
\frac{\Delta m~ m_2'}{m_1(m_1+m_2')} \Big[2 c^2 - \frac{(m_1\vec
u_1+m_2\vec u_2)^2}{(m_1+m_2)(m_1+m_2')}\Big]~~~,
\end{equation}
and $\hat v_{\rm out}$ is a kinematically free unit vector.  Denoting
by $\theta$ the angle between $\hat v_{\rm out}$ and the entrance channel
center of mass nucleon velocity $\vec u_1-(m_1 \vec u_1 +m_2 \vec u_2)/(m_1+m_2)=m_2(\vec u_1-\vec u_2)/(m_1+m_2)$,
and assuming that the center of mass scattering amplitude is a function $f(\theta)$ only of
this polar angle, the average over scattering angles of the outgoing nucleon velocity is given by
\begin{equation}\label{eq:vel2}
 \langle \vec v_1\rangle =\frac {m_1 \vec u_1 + m_2 \vec
u_2}{m_1+m_2'}+w \langle \cos \theta \rangle \frac{\vec u_1-\vec u_2}{|\vec u_1-\vec u_2|}~~~,
\end{equation}
 with $\langle \cos \theta \rangle$ given by
\begin{equation}\label{eq:costhetdef}
 \langle \cos \theta \rangle = \frac {\int_0^{\pi}d\theta \sin
\theta \cos \theta |f(\theta)|^2} {\int_0^{\pi}d\theta \sin \theta
|f(\theta)|^2}~~~.
\end{equation}
Subtracting $\vec u_1$ from Eq. \eqref{eq:vel2} gives the formula
for the average velocity change used in \cite{adler1} and \cite{adler2} to calculate the
flyby velocity change,
\begin{equation}\label{eq:vel3}
\langle \delta \vec v_1\rangle =\frac {m_2 \vec u_2 - m_2' \vec
u_1}{m_1+m_2'}+w\langle \cos \theta \rangle \frac {\vec u_1-\vec
u_2}{|\vec u_1-\vec u_2|}~~~.
\end{equation}

However, in addition to contributing to an average change in the outgoing nucleon velocity, dark
matter scattering will give rise to fluctuations in this velocity, which have a mean square magnitude
given by
\begin{equation}\label{eq:fluct1}
\langle (\vec v_1-\langle \vec v_1 \rangle)^2 \rangle
=w^2 \langle \Big(\hat v_{\rm out}-\langle \cos \theta \rangle  \frac{\vec u_1-\vec u_2}{|\vec u_1-\vec u_2|}\Big)^2 \rangle
=w^2 (1-\langle \cos \theta \rangle ^2)~~~.
\end{equation}
This fluctuating velocity leads to an average temperature increase of the nucleon, per single scattering, of
\begin{equation}\label{eq:tempincr1}
\langle \delta T \rangle= \frac {m_1}{2k_B} \langle (\vec v_1-\langle \vec v_1 \rangle)^2 \rangle
=\frac {m_1} {2k_B} w^2 (1-\langle \cos \theta \rangle ^2)~~~,
\end{equation}
with $k_B$ the Boltzmann constant.
In analogy with the treatment of the velocity change $\delta \vec v_1$ in  \cite{adler1}, to calculate $dT/dt$, the time rate of change of temperature
of the spacecraft resulting from dark
matter scatters,  one multiplies the
temperature change in a single scatter $\langle \delta T\rangle $ by the number of scatters per unit time.  This latter
is given by the flux $|\vec u_1-\vec u_2|$, times the scattering
cross section $\sigma$, times the dark matter spatial and velocity
distribution $\rho\big(\vec x, \vec u_2\big)$.  Integrating out
the dark matter velocity, one thus gets for $dT/dt$ at
the  point $\vec x(t)$ on the spacecraft trajectory with velocity
$\vec u_1=d\vec x(t)/dt$,
\begin{equation}\label{eq:tdot}
dT/dt= \int d^3 u_2 \langle \delta T\rangle |\vec
u_1-\vec u_2| \sigma \rho\big(\vec x, \vec u_2\big)~~~.
\end{equation}
Integrating  from
$t_i$ to $t_f$ we get for the temperature change resulting from
dark matter collisions over the corresponding interval
of the spacecraft trajectory ,
\begin{equation}\label{eq:tchange}
T_f-T_i  =\int_{t_i}^{t_f} dt \int d^3 u_2
 \langle \delta T\rangle |\vec u_1-\vec u_2| \sigma
\rho\big(\vec x, \vec u_2\big)~~~.
\end{equation}

In the elastic scattering case, with $\Delta m=0$, $m_2'=m_2$,
the formula of Eq. \eqref{eq:tdef} simplifies to
\begin{equation}\label{eq:el}
w^2 = \left( \frac{m_2}{m_1+m_2}\right)^2(\vec
u_1-\vec u_2)^2 ~~~.
\end{equation}
In the inelastic case, assuming that  $\Delta m/m_2$ and
$m_2'/m_2$ are both of order unity,  Eq. \eqref{eq:tdef} is well
approximated by
\begin{equation}\label{eq:inel}
w^2 \simeq    \Bigg( \frac{2 \Delta m ~m_2'} {
 m_1 (m_1+m_2')}\Bigg)c^2 ~~~.
\end{equation}
Since $\vec u_1$ and $\vec u_2$ are typically of order 10 ${\rm
km} ~{\rm s}^{-1}$, the temperature change in the inelastic case, per unit
scattering cross section times angular factors,  is
larger than that in the elastic case by a factor $ \sim c^2/|\vec
u_1|^2\sim 10^9.$

\section{Quick estimates}

Before going on to detailed modeling calculations using Eq. \eqref{eq:tchange}, we first
give quick estimates using Eqs. \eqref{eq:tempincr1},  \eqref{eq:el}, and \eqref{eq:inel},
making the approximations that the dark matter mass $m_2$ is much smaller than the nucleon
mass $m_1$, and that $\langle \cos \theta \rangle$ in Eq. \eqref{eq:tempincr1} is much smaller than 1.
In the elastic case, Eq. (4) of \cite{adler1} tells us that the magnitude of the
velocity change in a single collision is of order
\begin{equation}\label{eq:velche}
|\langle \delta \vec v_1 \rangle| \sim  \frac{m_2}{m_1} |\vec u_1-\vec u_2|~~~.
\end{equation}
Taking the ratio of the single collision temperature change to the single collision
velocity change, and multiplying by the flyby total velocity change $\sim 10^{-6}|\vec u_1|$,
we get as an estimate of the total temperature change
\begin{equation}\label{eq:este}
T_f-T_i \sim \frac{\delta T}{|\langle \delta \vec v_1 \rangle|}10^{-6}|\vec u_1|
\sim 10^{-6} \frac{m_2}{2k_B} |\vec u_1||\vec u_1-\vec u_2|
 \sim 0.6 \times 10^{-5}{}^{\circ}{\rm K} \Bigg(\frac{m_2 c^2}{ {\rm MeV}} \Bigg)~~~,
\end{equation}
in agreement with \cite{adler1}.

In the inelastic case, we must take into account the kinematic structure of an exothermic
inelastic differential cross section, as was done in \cite{adler2} (but was not correctly done in
the estimate given in \cite{adler1}).  In the inelastic case, Eq. (5) of \cite{adler1}
tells us that the magnitude of the velocity change in a single collision (for $\langle \cos \theta \rangle >0$) is of order
\begin{equation}\label{eq:velchi}
|\langle \delta \vec v_1 \rangle| \sim\frac{\sqrt{2 \Delta m \,m_2'}}{m_1} c \langle \cos \theta \rangle~~~.
\end{equation}
Writing the inelastic differential cross section near threshold in the form
\begin{equation}\label{eq:siginel}
\frac{d\sigma}{d\Omega}= \frac {A_{\rm inel}}{4\pi} k^{\prime} k^{-1} + B_{\rm inel}(k^{\prime})^2 \frac {3} {4\pi} \cos \theta + ...,
\end{equation}
we have
\begin{align}\label{eq:averages}
\sigma \simeq & A_{\rm inel}k^{\prime}  k^{-1}~~~,\cr
\langle \cos \theta \rangle \simeq& B_{\rm inel} k^{\prime} /(A_{\rm inel} k^{-1})~~~,\cr
\end{align}
with $k$ the entrance channel momentum
\begin{equation}\label{eq:kdef}
k=\frac{m_1 m_2}{m_1+m_2} |\vec u_1-\vec u_2| \simeq m_2 |\vec u_1-\vec u_2|~~~,
\end{equation}
and with $k^{\prime}$ the exit channel momentun, which to leading order in $\Delta m$ is 
\begin{equation}
k^{\prime}\simeq \sqrt{2 \Delta m \, m_2} c~~~.
\end{equation} 
Again taking the ratio of the single collision temperature change to the single collision
velocity change, and multiplying by the flyby total velocity change $\sim 10^{-6}|\vec u_1|$,
we get as an estimate of the total temperature change in the inelastic case
\begin{equation}\label{eq:esti1}
T_f-T_i \sim \frac{\delta T}{|\langle \delta \vec v_1 \rangle|}10^{-6}|\vec u_1|
\sim \frac{10^{-6}}{2k_B}\frac{A_{\rm inel}}{B_{\rm inel}} \frac{|\vec u_1|}{|\vec u_1-\vec u_2|}
\frac{1}{m_2}~~~.
\end{equation}
Defining a dimensionless parameter $S_{\rm inel}$ characterizing the inelastic
scattering by
\begin{equation}\label{eq:rdef}
\frac{A_{\rm inel}}{B_{\rm inel}} \equiv (m_2 c)^2 S_{\rm inel}~~~,
\end{equation}
the estimate of Eq. \eqref{eq:esti1} can be rewritten
as
\begin{equation}\label{eq:esti2}
T_f-T_i \sim 10^{-6} \frac{m_2 c^2}{2k_B} S_{\rm inel}\frac{|\vec u_1|}{|\vec u_1-\vec u_2|}
\sim 0.6 \times 10^4 {}^{\circ}{\rm K} S_{\rm inel} \Bigg(\frac{m_2 c^2}{ {\rm MeV}} \Bigg)~~~.
\end{equation}
Thus if $S_{\rm inel}$ is of order unity, the inelastic scattering temperature rise is substantially
bigger than that from elastic scattering, as already anticipated in the remarks following Eq. \eqref{eq:inel}
above.  For a dark matter-nucleon scattering force of range $a$, one expects $B_{\rm inel}/A_{\rm inel} \sim a^2$, 
giving the estimate $S_{\rm inel} \sim 1/(m_2^2 c^2 a^2) = (\lambda_2/a)^2$, with $\lambda_2=1/(m_2 c)$ the 
dark matter Compton wavelength.  So $S_{\rm inel}$ can be substantially less than unity.

Since $\langle \cos \theta \rangle \simeq k k^{\prime}/(S_{\rm inel}m_2^2 c^2)$, the approximation of neglecting
$\langle \cos \theta \rangle$ compared to 1 is equivalent to assuming that $S_{\rm inel} >> kk^{\prime}/(m_2 c)^2 = k k^{\prime} \lambda_2^2$. We continue to make this assumption in the numerical work of the next section, and note that it is compatible with having a small value of $S_{\rm inel}$.  

\section{Results from a model for the flyby velocity anomalies}

In \cite{adler2} we formulated a dark matter scattering model for the flyby velocity changes, by
assuming that inelastic and elastic dark matter scatterers populate shells generated by the precession
of circular orbits with normals tilted with respect to the earth's rotation axis.  By some simple
substitutions in the computer program used to calculate the velocity change predicted by a given
set of model parameters, one can calculate the corresponding temperature increase of the spacecraft
predicted by Eq. \eqref{eq:tchange}.  Combining the elastic and inelastic scattering contributions,
the  results are conveniently written in the form
\begin{equation}\label{eq:model}
 \frac{T_f-T_i}{{}^{\circ}{\rm K}} = \Bigg(\frac{m_2 c^2}{ {\rm MeV}} \Bigg) (Z_{\rm el} + S_{\rm inel} Z_{\rm inel})~~~,
\end{equation}
with $Z_{\rm el}$ and $Z_{\rm inel}$ dimensionless numbers giving respectively the elastic and inelastic
contributions to the temperature increase.  From fit 2d of \cite{adler2}, for which the elastic and inelastic shell radii
are 29,370 km and 34,520 km respectively, we give the fits to the velocity
anomalies and the corresponding values of $Z_{\rm el}$ and $Z_{\rm inel}$ in Table I.
Except for $Z_{\rm el}$ for the NEAR spacecraft, which in fit 2d barely intersects the elastic shell, the values of $Z_{\rm el}$ and $Z_{\rm inel}$
are in accord with the quick estimates made in the preceding section.

\section{Discussion and suggested experiment}

To summarize, if scattering from dark matter gravitationally bound to earth is responsible for the flyby
velocity anomalies, there must also be spacecraft temperature increases when the spacecraft passes
through the dark matter region.  This suggests two space science investigations.  The first is to analyze
the records of both earth-orbiting satellites and earth-exiting spacecraft, to see if there are unexplained
temperature anomalies, or at least to place bounds on temperature increases.  The second is to design a dedicated, compact,
thermally shielded  calorimetry experiment that could be carried as a secondary payload on future space missions,
to look for temperature increases as the spacecraft traverses the region within 70,000 km of the earth
associated with the flyby velocity anomalies.

\section{Acknowledgements}
This work was supported by the Department of Energy under grant DE-FG02-90ER40542.

\vfill\break

\begin{table} [t]\label{table:fits}
\caption{Flyby anomaly fit 2d and corresponding $Z_{\rm el}$ and $Z_{\rm inel}$ values.}
\centering
\begin{tabular} {|c| c |c| c| c| c |c|}
\hline\hline
~~~&~~~GLL-I~~~ & ~~~GLL-II~~~ & ~~~NEAR~~~ & ~~~Cassini~~~ & ~~~Rosetta~~~ & ~~~Messenger~~~ \\
\hline
$\delta v_{\rm A}$ (mm/s)  & 3.92 & -4.6 & 13.46 & -2 & 1.80 & 0.02 \\
$\sigma_{\rm A}$     (mm/s)  &0.3  & 1.0 & 0.01 & 1 & 0.03 & 0.01 \\
\hline
$\delta v_{\rm th}$   & 3.90    & -4.6& 13.46 & -2.7 & 1.80& 0.020\\
\hline
$Z_{\rm el}\times 10^5$& 0.26 & 0.85 & $0.17 \times 10^{-3}$ & 0.89 & 0.36 & 0.38 \\
$Z_{\rm inel}\times 10^{-4}$& 0.37 & 0.36 & 0.74 & 0.20 & 0.65 & 0.58 \\
\hline
\end{tabular}
\end{table}
\bigskip

The first two lines give the velocity discrepancy $\delta v_{\rm A}$ and the corresponding estimated error $\sigma_{\rm A}$
reported in \cite{anderson}.  The third line gives the theoretical values $\delta v_{\rm th}$ obtained from the model of
\cite{adler2}, which has two shells of dark matter, one containing elastic scatterers, the other containing inelastic
scatterers, gravitationally bound to the earth. The $Z_{\rm el}$ and $Z_{\rm inel}$ values, which give the flyby temperature
rise when substituted in Eq. \eqref{eq:model}, are given in the final two lines. (In fit 2d, on which this table is based,
the radius, Gaussian profile width, and tilt angle of the generating circular orbit are respectively 29,370 km, 6678 km, and 0.3902
radian for the elastically scattering shell, and 34,520 km, 3030 km, and 1.372 radian for the inelastically scattering shell.)

\end{document}